\documentclass[printer]{aa}

\usepackage{txfonts}
\usepackage[dvips]{graphicx}

\begin{document}

\title{Meteoroid and space debris impacts in grazing-incidence telescopes}

\author{J. D. Carpenter
\and A. Wells
\and A. F. Abbey
\and R. M. Ambrosi}

\institute{Space Research Centre, Department of Physics and Astronomy, University of Leicester, University Road, Leicester, LE1 7RH, UK\\
\email{james.carpenter@star.le.ac.uk}}
\date{February 25 2008}

\abstract
   {Micrometeoroid or space debris impacts have been observed in the focal planes of the XMM-Newton and Swift\--XRT (X-ray Telescope) X-ray observatories. These impacts have resulted in damage to, and in one case the failure of, focal-plane Charge-Coupled Device (CCDs) detectors.}
   { We aim to quantify the future risks of focal-plane impacts in present and future X-ray observatories.}
   {We present a simple model for the propagation of micrometeoroids and space debris particles into telescopes with grazing-incidence X-ray optics, which is based on the results of previous investigations into grazing-incidence hypervelocity impacts by microscopic particles. We then calculate micrometeoroid and space debris fluxes using the Micrometeoroid and Space Debris Terrestrial Environment Reference model (MASTER2005). The risks of future focal-plane impact events in three present (Swift\--XRT, XMM\--Newton, and Chandra) and two future (SIMBOL\--X and XEUS) X-ray observatories are then estimated on the basis of the calculated fluxes and the model for particle propagation.}
   {The probabilities of at least one impact occurring in the Swift\--XRT, XMM\--Newton, and Chandra focal planes, in a one year period from the time of writing in November 2007 are calculated to be $\sim5\%$ and $\sim50\%$ and $\sim3\%$. First-order predictions of the impact rates expected for the future SIMBOL\--X and XEUS X-ray observatories yield probablilities for at least one focal-plane impact, during nominal 5-year missions, of more than 94\% and 99\%, respectively.}
   {The propagation of micrometeoroids and space debris particles into the focal planes of X-ray telescopes is highest for Wolter optics with the largest collecting areas and the lowest grazing angles. Telescopes in low-Earth orbits encounter enhanced particle fluxes compared with those in higher orbits and a pointing avoidance strategy for certain directions can reduce the risk of impacts. Future X-ray observatories, with large collecting areas and long focal lengths, may experience much higher impact rates on their focal-plane detectors than those currently in operation. This should be considered in the design and planning of future missions.}

\keywords{X-ray telescopes, Swift, XMM, Chandra, meteoroid, space debris}

\titlerunning{Meteoroid and space debris impacts in grazing-incidence telescopes}
\authorrunning{J. D. Carpenter et al.}
\maketitle

\section{Introduction}

On 27 May 2005, after just 6-months in orbit, an anomalous event occurred in the focal-plane Metal Oxide Semi-conductor (MOS) Charge Coupled Device (CCD) detector of the Swift X-ray Telescope (XRT). This event has been interpreted as an impact by a micrometeoroid or space debris particle travelling at hypervelocity, such that the strength of the material is small compared with the inertial stresses during the impact. The event was observed as a bright light flash and the XRT went into its highest count-rate mode (Abbey \cite{abbey}). Post-event bright columns were observed in images, which were not previously present. The CCD focal plane of the XMM\--Newton X-ray telescope has experienced five similar events, one of which resulted in the failure of a CCD (Abbey et al. \cite{abbey}; Str\"uder et al. \cite{struder}). It is possible that particles, entering an X-ray telescope's optics at hypervelocity and impacting mirror surfaces at grazing-incidence angles, may be scattered into directions approximately parallel to the mirror surfaces and may enter the telescope via this process (Ambrosi et al. \cite{ambrosi}; Meidinger et al. \cite{meidinger}). Particles impacting with the focal-plane detectors, after entering a telescope, will produce craters and may degrade detector performance. In the case of Swift any future event may result in further degradation to the XRT or possibly the failure of the instrument if the single focal-plane CCD were to fail. For XMM\--Newton any future events will further degrade the performance of the telescope. 

Focal-plane micrometeoroid impact events are, to date, unique to XMM\--Newton and Swift\--XRT. Other similar grazing-incidence X-ray telescopes, such as Chandra, have not experienced similar events. In this paper we present a simple model for particle propagation and scattering in X-ray optics, and derive the micrometeoroid collecting power for Swift\--XRT, XMM\--Newton and Chandra. By applying a model for the fluxes of the micrometeoroid and space debris populations in various orbits, and scaling to the XMM\--Newton impact history, the probability of future impact events and the risks associated with various particulate sources have been estimated. From this analysis it can be shown that the probability of damage by impacts is dependent on a telecope's collecting area, optical geometry (bandpass), field of view and orbit. The impact rates for the planned SIMBOL\--X and XEUS X-ray telescopes are then predicted using the same model. It is shown that future X-ray telescopes with large collecting areas and long focal lengths will be at greater risk from impacting particles and should be designed with consideration for the effects of micrometeoroid impacts.

\section{Propagation of particles in X-ray telescopes}

It has been shown, experimentally (Ambrosi et al. \cite{ambrosi}; Meidinger et al. \cite{meidinger}) and with hydrocode models (Palmieri \cite{palmieri}) that microscopic particles incident at hypervelocity and at grazing-angles of up to and at least $5^{\circ}$, on highly polished mirror surfaces, are scattered into a velocity which is approximately parallel to the mirror. This occurs because the component of the projectile's velocity, which is normal to the mirror surface, is lost through plastic deformation of the projectile during the impact. It is possible therefore for meteoroid and space debris particles entering an X-ray telescope aperture at grazing incidence to be scattered from mirror shells and into a telescope's focal plane. 

\begin{figure*}
   \centering
	\includegraphics[width=12cm]{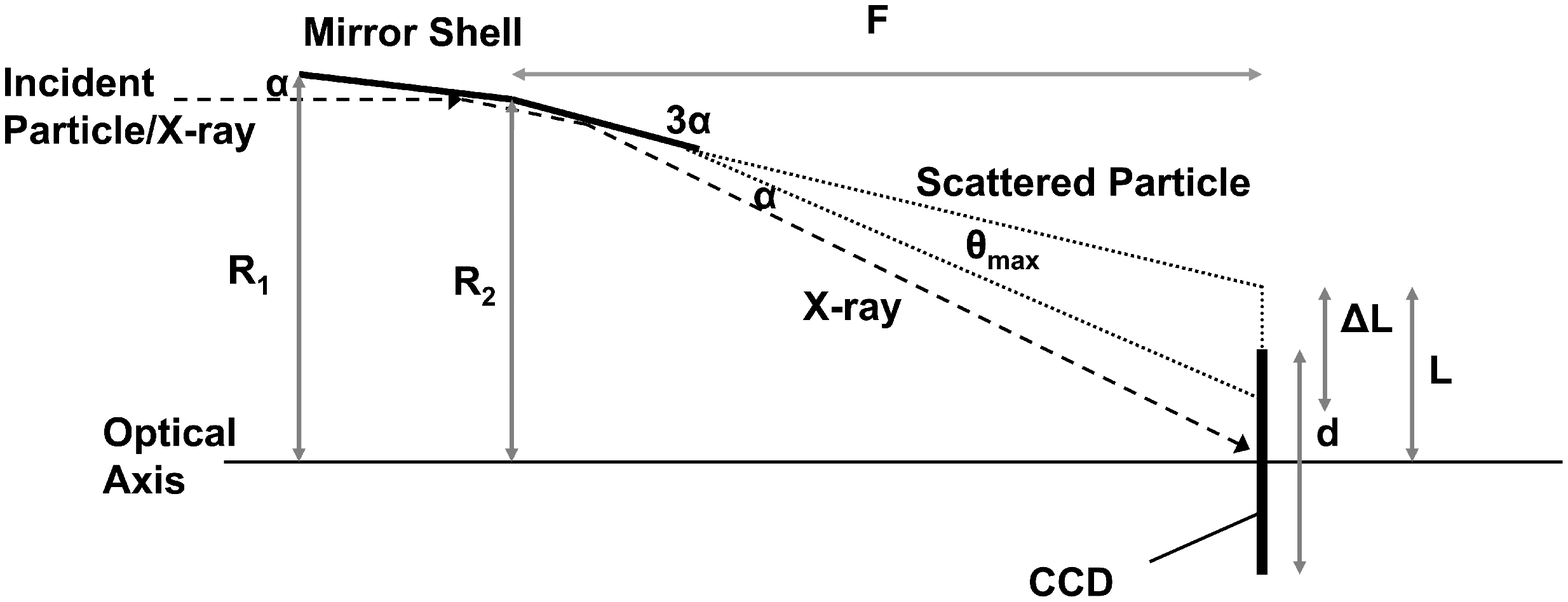}
   \caption{The path of a focussed X-ray and the range of possible paths for a meteoroid particle after incidence on a Wolter Type 
		I mirror assembly in an X-ray telecope.}
              \label{diagram}
    \end{figure*}

A simple model for particle propagation in a Wolter type I telescope assumes that all particles entering the telescope's aperture within a given acceptance cone, and which are subsequently incident on a mirror surface, are scattered into a direction approximately parallel to the mirror. A Wolter type I grazing incidence telescope is comprised of nested and sequential hyperboloid and paraboloid mirror shells (Van Speybroeck and Chase \cite{spey}). The angle at which a particle is incident upon the first mirror, measured from the mirror's surface, is denoted here as $\alpha$. After leaving the mirror in a direction, approximately parallel to its surface, the particle is incident at grazing incidence onto the second mirror, from which it is scattered a second time into some angle $\theta$ measured from the second mirror's surface. The path of a scattered particle into an X-ray telescope is shown in Fig. \ref{diagram}. Uncertainties in this model arise from the poorly defined scattering process and the range of incident angles for which this scattering process can be assumed. The model approximates the case for incident angles of up to and at least $5^{\circ}$ for which a gold mirror coating will not be perforated and little ejecta is likely to be produced by the impact (Palmieri \cite{palmieri}). No account is taken here of fragmentation by particles or the production of secondary ejecta particles from the mirrors. Observations in the Swift\--XRT and XMM\--Newton focal planes suggest multiple simultaneous impacts on the CCDs and so fragmentation or secondary-particle production processes are likely. This model also takes no account of particle composition. All modelled and experimental data to date have used spherical iron particles as projectiles and these are not representative of real micrometeoroids and space debris, which have complex structures and much lower densities, typically $2000 {\rm kg} {\rm m^{-3}}$ for the smallest cometary dust particles (Kessler \cite{kessler}) and $3500 {\rm kg} {\rm m^{-3}}$ for asteroidal dust particles (Staubach et al. \cite{staubach}).

Where $R_{1}$ is the front radius of the $n$th mirror shell's parabolic mirror and $R_{2}$ is the radius at the interface between the parabolic and hyperbolic  mirrors then the on-axis component of the $n$th parrabolic mirror area, $A_{n}$, is given by
\begin{equation}
A_{n}=\pi(R_{1}^2-R_{2}^2).
\end{equation}
If there is no straight through path to the rear hyperbolic mirrors, then the total on-axis geometrical area of a nest of $n$ mirrors is given by
\begin{equation}
A=\sum_{n}A_{n}.
\end{equation}
Assuming conical approximations to the paraboloid and hyperboloid mirror geometries for a telescope with focal length $f$, the grazing angle $\alpha_{n}$ between the optical axis and the $n$th parabolloid mirror shell is given by
\begin{equation}
\alpha_{n}=\arcsin(\frac{R_{1}}{4f}).
\end{equation}

For a grazing angle of $1^{\circ}$ a scattering probability, $P(\theta)$, distribution was observed by Meidinger et al. {\cite{meidinger}, in which the most likely scattering angle was $0^{\circ}$ from the surface, falling approximately linearly to a value $P(\theta)$ = 0 at an off axis angle $\theta_{max} =0.6^{\circ}$. In the following analysis a  probability distribution has been assumed where the probability of a particle being scattered through an angle $0^{\circ}<\theta<0.6^{\circ}$ is unity and the probability that a particle is scattered by an angle which is less than some angle $a$ is given by
\begin{equation}
P(0<\theta<a)=\frac{2(a-\frac{a^2}{2\theta_{max}})}{\theta_{max}}.
\end{equation} 

After scattering from the hyperboloid mirror a particle will exit at an angle from the mirror surface which is less than $\theta_{max}$. The conical approximation to the hyperboloid mirror subtends an angle $3\alpha$ to the optical axis. On axis X-rays, undergoing specular reflection at both mirrors, will emerge with an angle $4\alpha$ to the optical axis and will strike the focal plane at its centre. Micrometeoroid particles scattered from the mirror shells by an angle $\theta$ will propagate until ultimately arriving in the telescope's focal plane, where they may strike the detector. The detector area is approximated here as a circle, whose radius is determined by the telescope field of view and focal length. Particles are incident in the focal plane at a distance, $L$, from its centre, which is given by
\begin{equation}
L=f(\tan4\alpha-\tan(3\alpha+\theta)).
\end{equation}
The distance $\triangle L$ between the maximum and minimum possible impact positions in the focal plane is
\begin{equation}
\triangle L=f(\tan(3\alpha+\theta_{max})-\tan(3\alpha)).
\end{equation}
For a circular on-axis focal-plane detector with a diameter $d$, and a Field Of View (FOV) $\phi$
\begin{equation}
d=f\tan\phi.
\end{equation}
The length $\triangle d$ of the overlap of the scattered particle distribution in the focal plane and the detector is then given by 
\begin{equation}
\triangle d=\frac{d}{2}-L(\theta_{max})=f(\tan(3\alpha+\theta_{max})-\tan(3\alpha+a)).
\end{equation}
where $a$ is now the minimum scatter angle required to hit the detector so that a particle can impact upon the detector for angles greater than $a$. Equations 5 to 8 define the propogation of a particle in the model and the value of $a$. Using equation 4 the probability that a particle will strike the detector following a grazing-incidence impact is calculated as 
\begin{equation}
P_{hit}=1-P(0<\theta<a).
\end{equation}

The ``effective'' area to particles of a single mirror-shell pair is the product of the probability that an incident particle is scattered onto the detector and the on-axis geometric area of the shell. The effective area of the telescope is the sum of the effective areas to particles of all the nested mirror shells. An on-axis effective area, $A_{np}$, for a single mirror shell to micrometeoroids and debris particles is therefore
\begin{equation}
A_{np}=A_{n}P_{hit},
\end{equation}
and the total on-axis effective area to particles for the telescope $A_{p}$ is given by
\begin{equation}
A_{p}=\sum_{n}A_{np}.
\end{equation}

For particles incident at small off-axis angles a further factor of 0.5 is introduced because it is assumed that particles must be incident on the side of the optic with a smooth mirror surface in order to deflected into the focal plane. It may be however that scattering into the focal plane may also occur from the non-mirrored side of the optics. The effect of hypervelocity impacts on rough surfaces are unknown. In the following calculation of flux however this uncertainty is removed because impact event rates are calculated by scaling to that experienced by XMM\--Newton.

\begin{figure*}
   \centering
	\includegraphics[width=12cm]{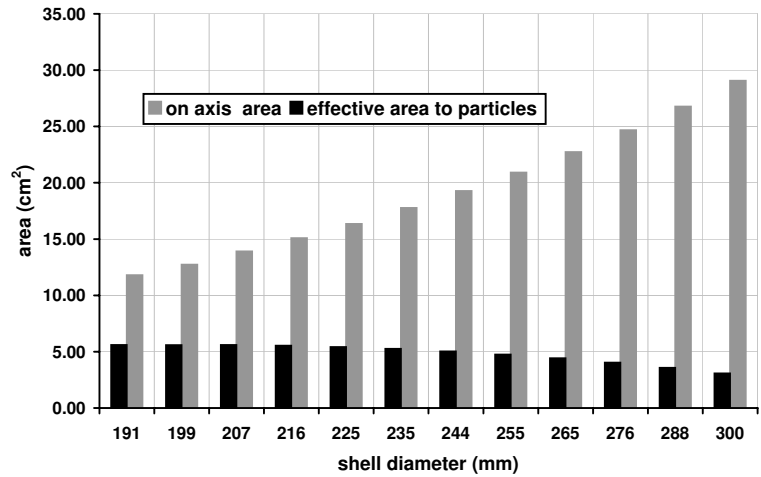}
   \caption{The on axis geometric areas and effective areas to meteoroid particles of the Swift\--XRT mirror shells.}
              \label{swift}
    \end{figure*}

\begin{figure*}
   \centering
	\includegraphics[width=12cm]{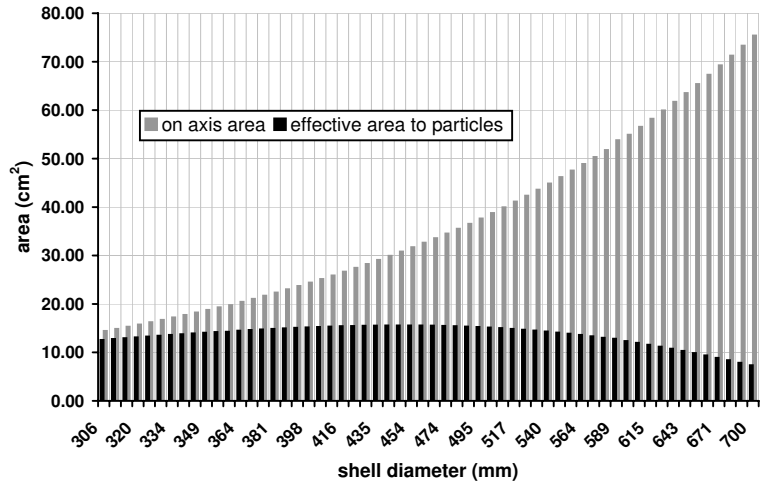}
   \caption{The on axis geometric areas and effective areas to meteoroid particles of the XMM\--Newton mirror shells.}
              \label{xmm}
    \end{figure*}

\begin{figure*}
   \centering
	\includegraphics[width=12cm]{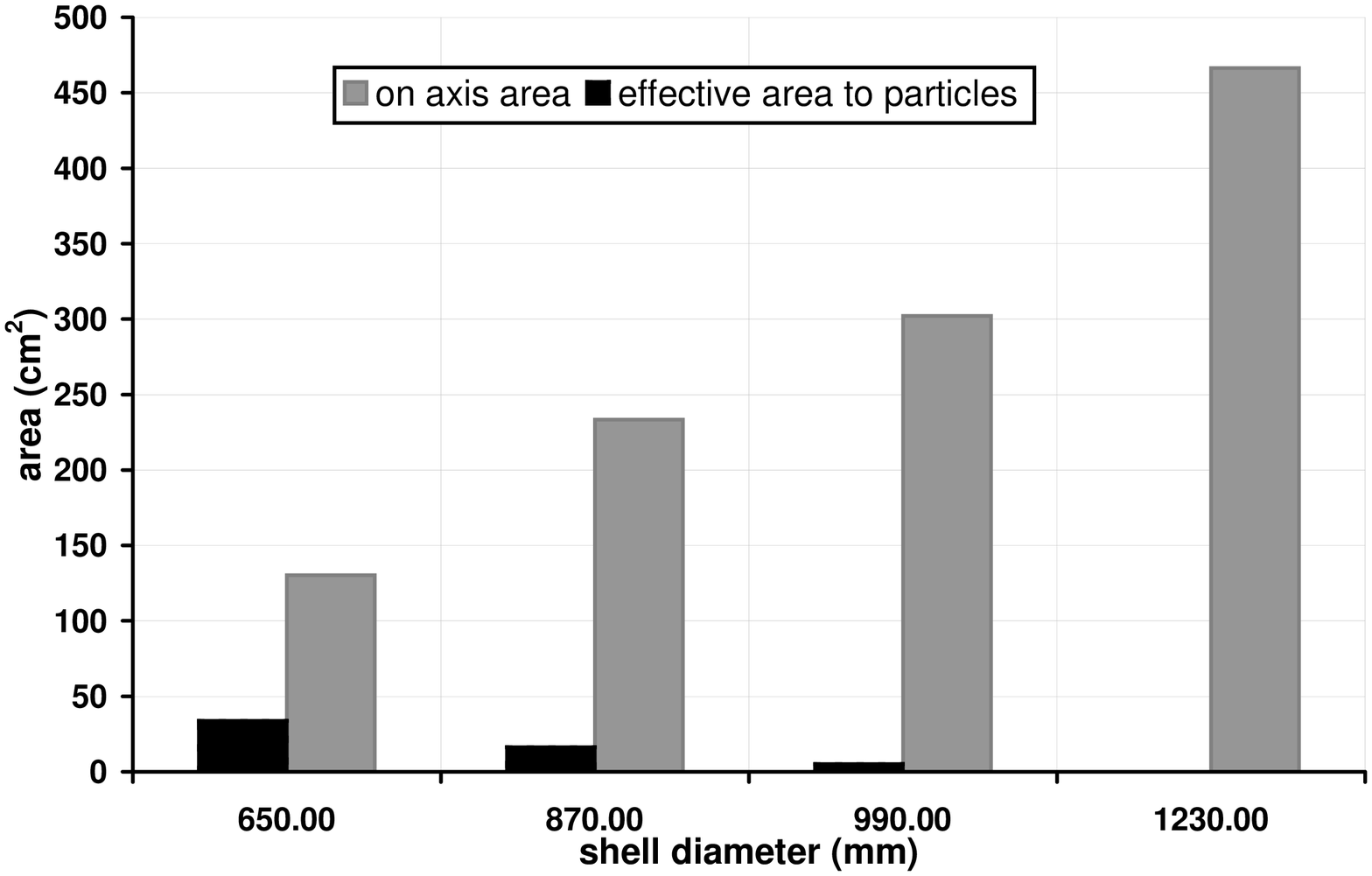}
   \caption{The on axis geometric areas and effective areas to meteoroid particles of the Chandra mirror shells.}
              \label{chandra}
    \end{figure*}

The effective area to incident particles as a function of mirror shell diameter, plotted along with the on axis geometrical collecting area is shown in Fig. \ref{swift} for Swift, Fig. \ref{xmm} for XMM-Newton and Fig. \ref{chandra} for Chandra. While the on axis geometric area and effective area to low energy X-rays of a nested Wolter optic is dominated by the largest mirror shells the primary contribution to $A_{p}$, and thus the flux of particles, is from the shells with the smallest diameters and grazing-angles. These mirror shells provide access to the highest X-ray energies but also provide access to meteoroid populations. The focal-plane micrometeoroid impact rate is therefore a function of both a telescope's geometric collecting area and its X-ray bandpass. The total on-axis effective area to particles of Swift\--XRT is calculated to be $58 {\rm cm^{2}}$. For XMM-Newton two (one and two halves) of the three separate telescopes have CCD focal planes. The effective area to particles for two XMM\--Newton telescopes is calculated to be $1574 {\rm cm^{2}}$. The XMM\--Newton and Chandra X-ray telescopes have similar high Earth orbits and will therefore experience similar micrometeoroid fluxes, and yet Chandra has experienced no impact events to date. The comparatively large numbers of mirror shells with small radii and shallow grazing-angles on XMM\--Newton lead to a larger effective area to particles than that of Chandra, for which the effective area is calculated to be  $55 {\rm cm^{2}}$. The Chandra focal-plane CCD impact rate is therefore estimated to be approximately $\frac{1}{29}$ that experienced by XMM\--Newton. 

\section{Micrometeoroid and debris fluxes}

The fluxes of micrometeoroid and space debris particles have been simulated using the Micrometeoroid And Space debris Terrestrial Environment Reference model (MASTER2005) (Oswald et al. \cite{osw}). This model is based on the combination of a semi-deterministic analysis of a reference population, derived from the major components of the space debris population, and natural meteoroid fluxes determined using the Divine-Staubach meteoroid model (Divine et al. \cite{divine}). Fluxes due to the meteor streams are modelled using the McBride/Jenniskens meteoroid steams model (Jenniskens \cite{jeniskens}; McBride \cite{mcbride}). Anthropogenic debris fluxes considered by the model include spent payloads and upper stages, fragments following on-obit collisions and explosions, aluminium oxide dust particles and slag from solid rocket motor (SRM) burns, sodium-potassium (NaK) nuclear reactor coolant expelled from Soviet RORSAT satellites, particles released from degrading spacecraft surfaces (referred to as paint flakes) and ejecta from impacts on spacecraft surfaces. 

The incident particle flux as a function of particle type, relative velocity, diameter, and incident direction with respect to the Swift velocity vector was determined. The fluxes simulated here are for particles with diameters greater than $1 \mu {\rm m}$ and are the mean fluxes for the period between launch in November 2004 and the 1 May 2005, after which MASTER2005 cannot determine debris fluxes for particles with diameters less than 1 mm.  

The flux of natural interplanetary dust particles (IDPs) is enhanced in low-Earth orbit by a factor of $\sim$2, compared with interplanetary space, by gravitational focussing (McDonnell et al. \cite{mcdonnell}). Debris fluxes can be highly directional as they tend to result from particles with orbits related to those of their parent bodies. Strong peaks in the flux of debris particles are associated with orbital inclinations of approximately $75^{\circ}$, $63^{\circ}$, $82^{\circ}$, $100^{\circ}$ (Sun synchronous orbits), $90^{\circ}$ (polar orbits), $0^{\circ}$ (Geosynchronous Earth Orbits (GEO)) and $28.5^{\circ}$ (launches going due East from Kennedy Space Centre)(McDonnell et al \cite{mcdonnell}). The particle fluxes experienced by spacecraft in high-Earth orbits, above GEO, cannot be predicted using MASTER2005 but a good approximation is obtained by modelling only the natural component of the flux (the Divine-Staubach flux) for a spacecraft in GEO and thus minimising gravitational enhancement (Oswald \cite{oswa}). The debris component of the particle flux in high-Earth orbits can be assumed to be negligible compared with the flux of IDPs.

The total flux of particles with diameters $\geq 1 \mu {\rm m}$ incident on a sphere in a Swift-like orbit (semi major axis = $6978.14 {\rm km}$, eccentricity = 0.0015, inclination = $20.56^{\circ}$, right ascension of ascending node = $358.17^{\circ}$, argument of perigee = $68.47^{\circ}$) is calculated to be $5383 {\rm m^{-2}} {\rm yr^{-1}}$, accounting for the Swift pointing strategy in which the telescope never points within $27^{\circ}$ of the Earth's limb or within $7 ^{\circ}$ of the velocity vector. The mean relative velocity at impact is calculated to be $15 {\rm km} {\rm s^{-1}}$. The flux as a function of particle diameter for a sphere in a Swift-like orbit is shown in Fig. \ref{flux}. For XMM\--Newton and Chandra in high-Earth orbits the flux is calculated to be $2957 {\rm m^{-2}} {\rm yr^{-1}}$. The contributions to the total fluxes encountered by XMM-Newton and Chandra at perigee, near the Earth, are small. 

\begin{figure*}
   \centering
	\includegraphics[width=12cm]{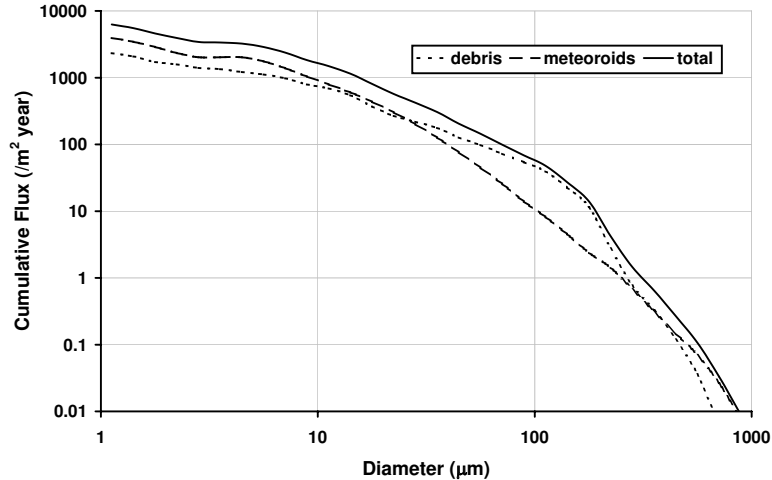}
   \caption{The mean flux, as a function of particle diameter, of meteoroids and space debris particles incident on the Swift spacecraft calculated using the MASTER2005 model.}
              \label{flux}
    \end{figure*}

The contribution to the mean flux calculated here from the seasonal meteoroid streams is small compared with that due to other particle sources, however due to their highly directional and seasonal properties stream particles may generate significant increases in flux over short periods of time, associated with specific pointing directions. Streams may therefore pose a hazard to X-ray telescopes, which is associated with observations of particular targets at particular times of the year and meteoroid stream avoidance strategies are already applied in the operations of some X-ray telescopes. Stream particles tend to have velocities as high as $80 {\rm km} {\rm s^{-1}}$, which are much greater than for background IDPs (typically $8-20 {\rm km} {\rm s^{-1}}$). The detection threshold (essentially a measure of the damage caused by an impact) for incident particles by dust detectors has a typical power law dependence on velocity $V^{\gamma}$, where $\gamma$ can be as high as $\sim3.5$ (McDonnell \cite{mcdonnell}). Meteor-stream particles may therefore be as much as three orders of magnitude more ``detectable'' by an X-ray telescope than the micrometeoroid background. 

Parameters used to describe particle fluxes associated with 50 meteoroid streams have been determined, based on data from ground based observations of meteor showers, by Jenniskens (\cite{jeniskens}) and McBride (\cite{mcbride}). The flux Vs. particle mass distributions for these streams are strictly only applicable to the minimum masses associated with the faintest visible meteors. A working mass limit of $10^{-10} {\rm kg}$ is generally assumed, pertaining to faint radio meteors. Assuming a particle density typical for cometary IDPs of $2000 {\rm kg} {\rm m^{-3}}$, this lower mass limit corresponds to a particle diameter of $\sim50 \mu {\rm m}$. Smaller particles in the streams will have short lifetimes as they are removed by radiation pressure and become part of the background IDP flux. The cumulative flux of particles with diameters greater than $50 \mu {\rm m}$ is calculated to always be less than $10^{-6} {\rm m^{-2}} {\rm s^{-1}}$ for any meteor stream. 

\section{Focal-plane event rates and impact probabilities}

It has been shown by previous authors (Ambrosi et al. \cite{ambrosi}; Meidinger et al. \cite{meidinger}; Palmieri \cite{palmieri}) that, for incidence angles up to at least $5^{\circ}$, particles incident on X-ray mirrors will be scattered but the upper limit on the incidence angle required for scattering of particles is unknown. This provides an uncertainty in determinning the acceptance cone for particles into a grazing-incidence telescope. Here we estimate the acceptance angle upper limit by scaling the cone angle such that the fluxes calculated for XMM-Newton are in agreement with the observed number of 5 impact events in 7.75 years of operation (at time of writing), giving a mean impact rate of 0.65 focal-plane events per year. As is common for dedicated in-situ measurements of interplanetary dust, the small number of detected events leads to large uncertainties in the mean rate. For a measured sample of 5 events the true mean number of events is between 1.97 and 10.51 to a confidence level of 0.9 (Gr\"un \cite{grun}) giving a range of possible mean event rates between 0.25 and 1.4 events per year. For acceptance cone half-angles of $7.5^{\circ}$, $8^{\circ}$ and $8.5^{\circ}$ event rates of 0.55, 0.71 and 0.80 events per year are predicted by the model. On this basis the acceptance angle for particles into the Wolter optics is expected to be $\sim8^{\circ}$. Fig. \ref{directions} shows the calculated mean flux of particles incident within $8^{\circ}$ of the Swift\--XRT's optical axis as a function of pointing direction, where the pointing direction is measured locally with respect to Swift's velocity vector. The major contribution to the flux is from natural meteoroids but there is a highly directional flux of debris particles, shown as two bright spots in Fig. \ref{directions}. These debris particles derive from parent bodies in polar orbits. There is a general enhancement in flux for pointing directions which have a positive ram component (i.e. a velocity component parallel to that of the spacecraft). The Swift operational strategy is to not point within $7^{\circ}$ of the ram direction but from Fig. \ref{directions}, and assuming an $8^{\circ}$ acceptance angle for particles, it is evident that this is of little benefit in terms of reducing the risk of impacts. Another Swift operational strategy is to not point within $27^{\circ}$ of the Earth's limb. This strategy acts to exclude the majority of the highly directional debris particles in near polar orbits.

\begin{figure*}
   \centering
	\includegraphics[width=12cm]{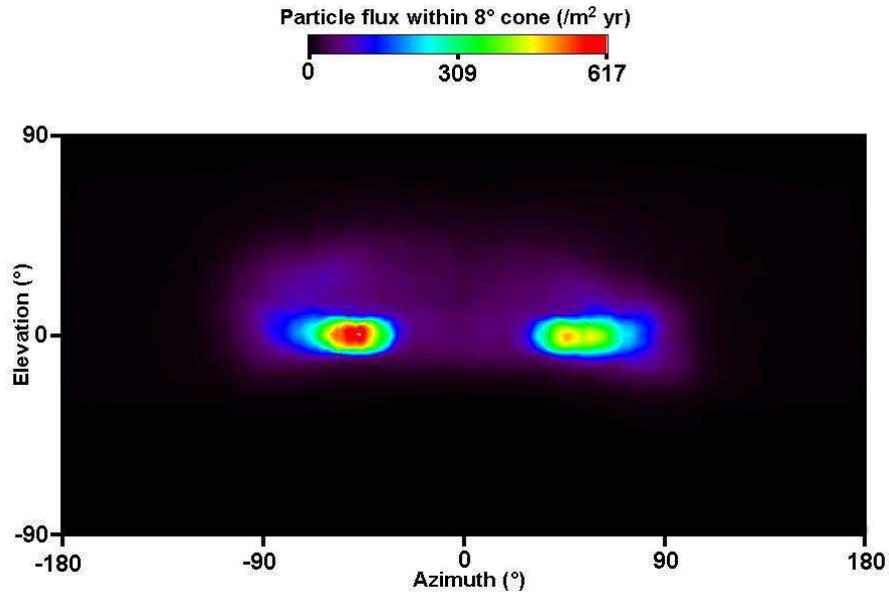}
   \caption{The calculated flux of meteoroids and debris particles incident within $8^{\circ}$ of the Swift\--XRT's optical axis as a function of pointing direction. The pointing direction of the telescope is measured in degrees from the the velocity vector of the spacecraft.}
              \label{directions}
    \end{figure*}

If it is assumed that the acceptance angle to particles is the same for all X-ray telescopes then the event rates for any X-ray telescope can be estimated by scaling to the observed XMM-Newton event-rate. The event-rate $N_{p}$ in a telecope with effective area to particles $A_{p}$ and which experiances a mean flux of particles on its aperture $F_{p}$ is given by 
\begin{equation}
N_{p}=N_{xmm}\frac{A_{p}F_{p}}{A_{xmm}F_{xmm}},
\end{equation}
where $N_{xmm}$, $A_{xmm}$ and $F_{xmm}$ are the event rate, effective area to particles and particle flux experienced by XMM\--Newton.

The mean focal-plane event rate predicted for the Swift\--XRT is calculated to be between 0.02 and 0.09 per year to a confidence level of 0.9. The probability of the observed one impact event during the first 2.75 years of operation is between 4\% and 19\%. In the coming year the probability of at least one impact occurring is between 1.7\% and 8.6\%. 

The predicted mean focal-plane event rate for the Chandra is calculated to be between 0.01 and 0.05 per year to a confidence level of 0.9. The probability of no impacts in 8.17 years of operation is between 66\% and 92\%. In the coming year the probability of at least one impact occurring is between 1\% and 5\%. 

The probability of an impact occurring in the XMM\--Newton focal plane in the coming year is between 22\% and 75\%. 

For meteoroid streams the predicted focal-plane event rate for XMM-Newton is always less than $2\times10^{-7} {\rm s^{-1}}$ assuming an on-axis observation of the stream and zero divergence of the stream from the radiant direction. Thus for all current observatories there is little benefit in imposing a stream-avoidance strategy to reduce the risk of impacts during streams.

\section{Future X-ray telescopes}
In this section we estimate the micrometeoroid impact rates for two future X-ray telescopes; SIMBOL\--X (Ferrando et al. \cite{ferrando}) and XEUS, using approximations to their optical geometries. The calculated values are first order predictions for impact rates in the telescopes' focal planes.

SIMBOL\--X is modelled here as a telescope with a 30m focal length and a 6 arcminute FOV. The optics are approximated as 100 nested mirror shells, which are conical approximations to the true Wolter Geometry. The external diameters of the input mirrors vary between 0.29m and 0.6m. The micrometeoroid fluxes experienced by SIMBOL\--X are assumed to be identical to those for XMM\--Newton and Chandra. The loss of collecting area due to the thickness of the individual mirror shells is ignored. The effective area and geometric area for collection of each of the modelled SIMBOL\--X mirror shells is shown in Fig. \ref{simbolx}. The total calculated on-axis effective area to particles for the SIMBOL\--X telescope is $1700 {\rm cm^{2}}$. The predicted mean focal-plane event rate for SIMBOL\--X is calculated to be between 0.56 and 2.9 per year to a confidence level of 0.9. The probability of at least one impact occurring during 5 years of operation is greater than 94\%.  

\begin{figure*}
   \centering
	\includegraphics[width=12cm]{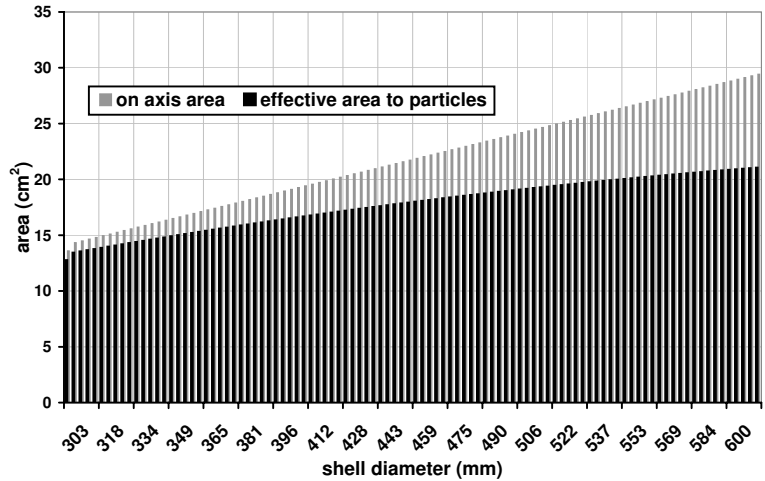}
   \caption{The on axis geometric areas and effective areas to meteoroid particles predicted for the SIMBOL-X mirror shells.}
              \label{simbolx}
    \end{figure*}

The XEUS optics are approximated here as 500 nested mirror shells with diameters ranging between 0.66m and 2.2m. The focal length of the telescope is 35m and the telescope's FOV is 7 arcminutes. The micrometeoroid fluxes experienced by XEUS are assumed to be identical to those for XMM\--Newton and Chandra. The loss of collecting area due to the thickness of the individual mirror shells is ignored. The effective area and geometric area fore collection of each of the modelled XEUS mirror shells is shown in Fig. \ref{xeus}. The total on-axis effective area to particles of the XEUS telescope is calculated to be $1.2 {\rm m^{2}}$. The predicted mean focal-plane impact event rate for XEUS is calculated to be between 3.4 and 20 per year to a confidence level of 0.9. The probability of at least one impact occurring during 5 years of operation is greater than 99\%.

\begin{figure*}
   \centering
	\includegraphics[width=12cm]{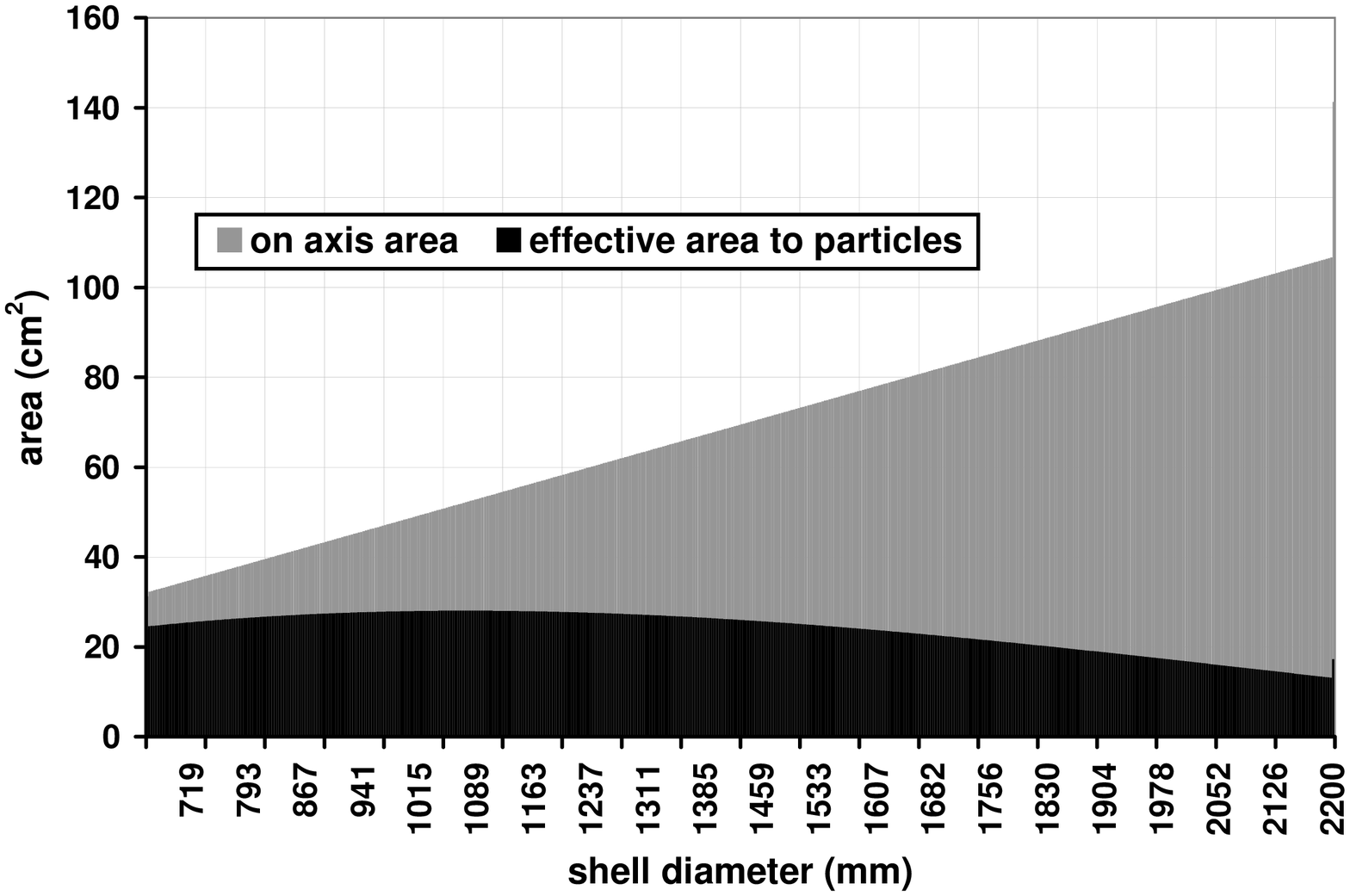}
   \caption{The on axis geometric areas and effective areas to meteoroid particles predicted for the XEUS mirror shells.}
              \label{xeus}
    \end{figure*}

\section{Summary and conclusions}
Small micrometeoroids and space debris particles, incident at shallow grazing-angles on grazing-incidence Wolter optics, are scattered into directions approximately parallel to the mirror surfaces and can be incident on detectors in a telescope's focal plane. Particle impacts on CCDs can damage individual pixels or columns and can ultimately lead to detector failure. Such impacts are the probable cause of anomalous events and subsequent damage observed to the focal-plane CCD detectors of the XMM\--Newton and Swift\--XRT X-ray telescopes. 

The particle flux incident on a detector will be dominated by particles scattered from mirror shells, for which grazing angles are small. Thus X-ray telescopes with large effective areas at the highest X-ray energies will be most vulnerable to impacts on focal-plane detectors. Telescopes in low-Earth orbit experience particle fluxes which are enhanced compared with those in higher orbits by gravitational focussing of natural meteoroids and by space debris particles, which tend to be incident from well defined directions. Pointing strategies which avoid these key directions may be beneficial to the operations of Swift and other X-ray telescopes in low-Earth orbits. 

A range of possible impact rates and probabilities for future impact rates have been predicted for three currently operating X-ray observatories. It is shown that the optical geometry on XMM-Newton is favourable to the propagation of particles into the focal plane, whereas for Chandra increased grazing angles make focal-plane impacts less likely. Impacts in the Swift\--XRT focal plane are unlikely events, although the increased particulate fluxes in low Earth orbit lead to an increased impact risk. 

The risk of impacts on two future observatories has also been estimated. Future X-ray telescope collecting areas will be significantly larger than those of exisiting observatories and focal lengths will be much greater. These changes result in an increase in the overall collecting area for meteoroids and therefore in the focal-plane impact rate. In addition increased collecting area also increases the upper size limit of the particles being sampled. As a result the risk of damage to focal-plane instruments, by particle impacts, will increase. It is therefore essential that micrometeoroid and debris impacts are considered as part of the design and mission planning of future X-ray telescopes.

Future work should investigate grazing-incidence hypervelocity impacts on X-ray mirror surfaces, observing the effects of incident angle and velocity on acceptance angles, scattering angles and particle fragmentation. This will allow the development of a more detailed model for the assessment of the risks posed by meteoroids and debris particles to grazing-incidence observatories. An understanding of the relation between the properties of impacting particles and the failure and degradation of CCDs is also essential for determining risks to future and current missions. 

\begin{acknowledgements}
The authors wish to acknowledge the following, Julian Osbourne of the University of Leicester and the Swift XRS team for supporting the study, Michael Oswald of the Technical University of Braunschweig for providing support for the MASTER2005 modelling, Neil McBride for advice in the interpretation of meteoroid stream fluxes and Gerhard Drolshagen of the European Space Agency for providing technical information and advice. 
\end{acknowledgements}

\end{document}